\documentclass[prb, reprint]{revtex4-2}

\pdfoutput=1
\usepackage{amsmath, amssymb, graphicx}
\usepackage[utf8]{inputenc}
\usepackage[english]{babel}
\usepackage{times, txfonts}
\usepackage{hyperref}

\newcommand{\bk}{\mathbf{k}}
\newcommand{\br}{\mathbf{r}}
\newcommand{\psio}{\psi_{\bk \, {=} \, 0}}

\renewcommand{\Im}{\mathop{\mathrm{Im}}}

\begin{document}

\date{April 26, 2021}

\title{Loop parametric scattering of cavity polaritons}

\author{S.~S.~Gavrilov}

\affiliation{Institute of Solid State Physics RAS, 142432 Chernogolovka, Russia}
\affiliation{National Research University Higher School of Economics, 101000 Moscow, Russia}

\begin{abstract}
  Within the framework of the mean-field approximation, a coherently
  excited two-dimensional system of weakly repulsive bosons is
  predicted to show a giant loop scattering when the rotational
  symmetry is reduced.  The considered process combines (i) the
  parametric decay of the driven condensate into different $k$-states
  and (ii) their massive back scattering owing to spontaneous
  synchronization of several four-wave mixing channels.  The
  hybridization of the direct and inverse scattering processes, which
  are different and thus do not balance each other, makes the
  condensate oscillate under constant one-mode excitation.  In
  particular, the amplitude of a polariton condensate excited by a
  resonant electromagnetic wave in a uniform polygonal GaAs-based
  microcavity is expected to oscillate in the sub-THz frequency
  domain.
\end{abstract}

\maketitle

\section{Introduction}
\label{sec:intro}

Two-dimensional cavity polaritons are a result of exciton-photon
coupling in layered
heterostructures~\cite{Weisbuch1992,Yamamoto.book,Kavokin.book.2017}.
Being composite bosons, they exhibit two kinds of coherent states, one
of which is similar to Bose-Einstein condensates (BECs) formed with
decreasing temperature~\cite{Kasprzak2006}, whereas the other appears
when a resonant electromagnetic wave excites polaritons
directly~\cite{Baas2006}.  Both kinds of coherent states are
characterized by a mean-field amplitude $\psi(\br, t)$ obeying a
generalized wave equation
\begin{equation}
  \label{eq:gp}
  i \hbar \frac{\partial \psi}{\partial t} = \left[ E(\mathbf r,
    -i \hbar \nabla) - i \gamma + V \psi^* \psi \right]
  \psi + f(\br, t)
\end{equation}
(spin/polarization degrees of freedom are disregarded).  If the
pumping force $f$ and decay rate $\gamma$ are zero, Eq.~(\ref{eq:gp})
is reduced to the Gross-Pitaevskii equation for equilibrium BECs.
Similar to atomic gases, cavity polaritons combine repulsive
interaction ($V > 0$) and positive mass in the vicinity of the
ground-state level
$E_g = E(\bk \, {=} \, 0)$~\cite{Yamamoto.book,Kavokin.book.2017}.

Under plane-wave pumping
$\bigl[ \text{e.\,g.,}~f(\br, t) = \bar f e^{i (\bk_p \br - E_p t /
  \hbar)} \bigr]$, the condensate has the same wave vector and
frequency as the pump wave, provided that $\gamma > 0$ and $E_p$ is
not too far from resonance.  The forced oscillation of $\psi$ results
in deep qualitative changes of the Bogolyubov excitation spectrum
$\tilde E(\bk)$ compared to equilibrium
systems~\cite{Ciuti2003,Gippius2004.epl}.  In particular, the
excitations around $\bk_p = 0$ are no longer sonic unless $E_p$ is
equal to $E_g + V|\psio(\bar f)|^2$ for a given pump amplitude
$\bar f$~\cite{Carusotto2013}.  Besides, as $\bar f$ and
$|\psi_{\bk_p}(\bar f)|$ are increased, the sign of $\Im \tilde E$ may
reverse at some $\bk = \bk'$, which means the instability of the
condensate against two-particle scattering
$(\bk_p, \bk_p) \to (\bk', 2 \bk_p - \bk')$, often leading to a strong
redistribution of polaritons in the $\bk$ space.  For instance, the
break-up of the condensate excited with a nonzero $\bk_p$ near the
inflection point of $E(|\bk|)$ is known to result in macroscopic
occupation of two modes $\bk \approx 0$ and
$\bk \approx 2 \bk_p$~\cite{Savvidis2000,Stevenson2000,Butte2003}.
Such processes, which attracted much interest in the early 2000's,
were firstly understood by analogy with optical parametric oscillators
(OPOs) \cite{Whittaker2001,Ciuti2001} in which an external pump beam
splits into a pair of plane waves, conventionally referred to as
``signal'' and ``idler''.  This analogy is not perfect because the
polariton OPO arises through fluctuations~\cite{Dagvadorj2015} and
never comes to a state with only three nonempty wave
modes~\cite{Savvidis2001,Whittaker2005,Gavrilov2007.en}.
Nevertheless, the condensate induced by coherent pumping usually
remains the most populated mode that governs all signals and idlers
excited owing to the parametric scattering~\cite{Demenev2008,
  Krizhanovskii2008, Wouters2007.prb.threshold, Dunnett2018}.

Here, we report an unusual manifestation of the parametric scattering,
which is expected to occur in wide (tens of $\mu$m) and spatially
uniform samples of a polygonal shape.  We show that a reduced
rotational symmetry leads to a sort of population inversion such that
a number of scattered modes get noticeably stronger than the pumped
mode ($\bk_p = 0$, $E_p > E_g$).  Consequently, new two-particle
interaction processes come into play which are different from the
direct break-up of a pumped condensate into signals and idlers;
furthermore, the pumped mode can itself act as a parametric signal.
By virtue of symmetry, several processes of such kind synchronize,
share the same target state $\bk = 0$, and thus yield a massive back
scattering of polaritons.  It is important that the direct and inverse
scattering effects do not cancel each other but constitute a unified
loop interaction process.  The common signal of the back scattering
arises near the ground-state energy level $E_g$ rather than at the
pump level $E_p$.  As a result, the pumped mode has two energy peaks
and its amplitude $|\psio|$ oscillates at frequency
${\sim} \, (E_p - E_g) / \hbar$.  At the same time, several scattered
modes with $\bk \neq 0$ remain nearly steady and very strong, being
thus a dynamical reservoir that feeds the new condensate.

In what follows, all these phenomena are considered in detail.  We
begin in Sec.~\ref{sec:blowup} with describing the feedback between
the externally pumped and scattered polariton modes which is
responsible for their abnormal population.  The same feedback
mechanism was found earlier in isotropic systems, both
infinite~\cite{Gavrilov2014.prb.b,Gavrilov2015} and strongly
confined~\cite{Whittaker2017}, where it had different observable
manifestations.  In Sec.~\ref{sec:loop} we demonstrate the onset of
the macroscopic loop interaction in a square cavity.  Finally, in
Sec.~\ref{sec:discussion}, we summarize the results and compare them
to some recent studies dealing with the dynamical
condensation~\cite{Sun2012} and self-pulsations in a polariton fluid
under coherent driving~\cite{Gavrilov2016,Leblanc2020}.

\section{Blowup and abnormal population of scattered modes}
\label{sec:blowup}

\begin{figure}
  \centering
  \includegraphics[width=\linewidth]{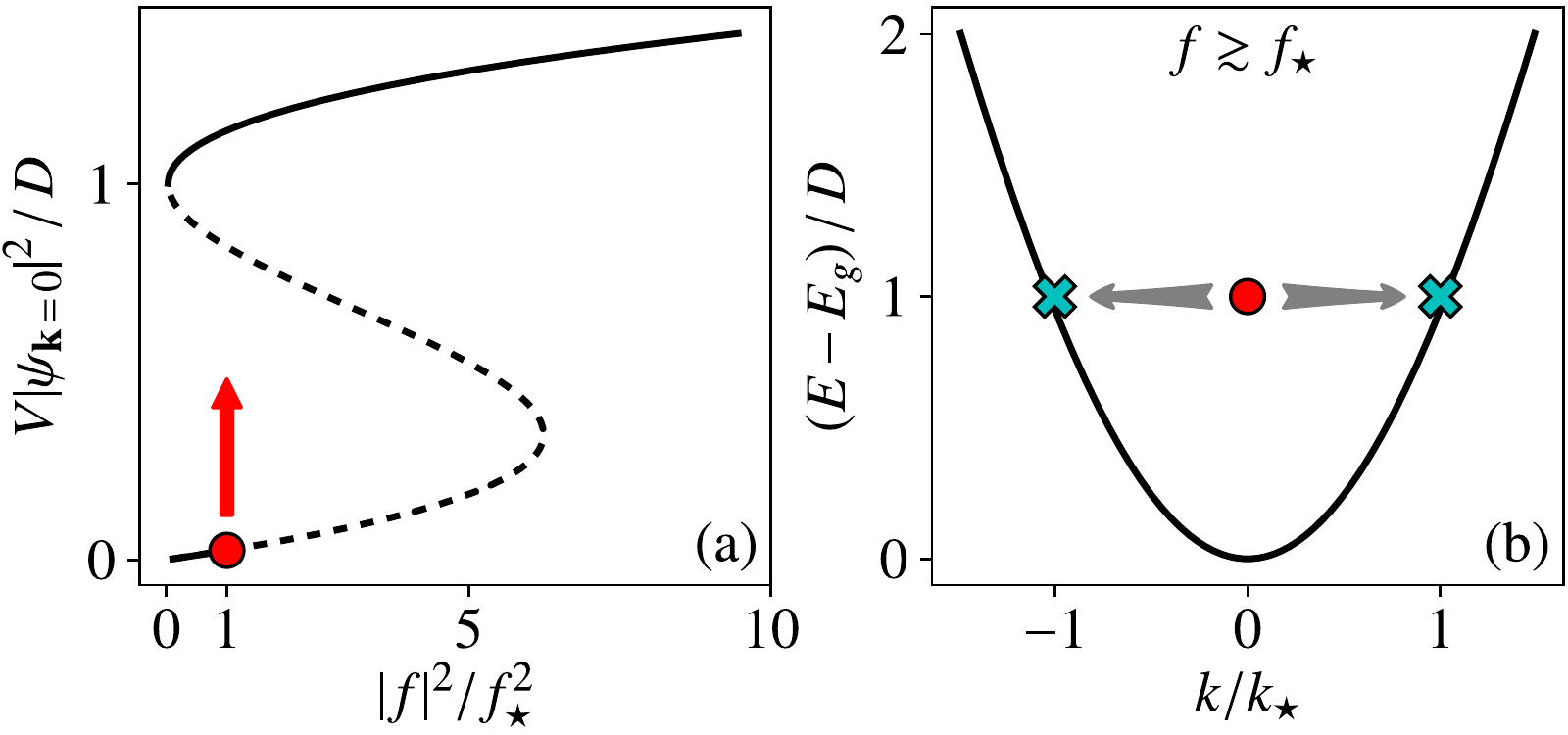}
  \caption{\label{fig:s} (a) One-mode response of a driven condensate;
    dashed line indicates unstable solutions;
    $D = E_p - E_g = 40 \gamma$ is the pump detuning from the
    resonance at $\bk_p = 0$;
    $f_\star = \sqrt{(\gamma / V) \bigl[ (D - \gamma)^2 + \gamma^2
      \bigr]}$ \cite{Gavrilov2014.prb.b}. (b)~Dispersion law, position
    of the pumped mode, and scheme of the parametric scattering. The
    scattering wave number $k_\star \approx 0.9~\mu\mathrm{m}^{-1}$.}
\end{figure}

Let us recall several basic facts about an interplay between the
parametric scattering and bistability in a homogeneous and isotropic
polariton system.  It is known that the one-mode dependence of
$|\psi_{\bk_p}|^2$ on $|f|^2$ has an `S'-shaped form
[Fig.~\ref{fig:s}(a)] as long as
$D \equiv E_p - E(\bk_p) > \sqrt{3} \gamma$~\cite{Elesin1973,
  Baas2004.pra, Gippius2004.epl, Carusotto2004}.  When
$|\psi_{\bk_p}|$ is small, the repulsive interaction of polaritons
($V > 0$) involves a blueshift of their resonance energy
$E(\bk_p) + V |\psi_{\bk_p}|^2$ towards the pump level $E_p$,
resulting in a superlinear increase of $|\psi_{\bk_p}|$ as a function
of $|f|$ throughout the lower branch of solutions.  On the upper
branch, by contrast, the dependence of $|\psi_{\bk_p}|$ on $|f|$ is
sublinear, because the effective resonance energy has exceeded $E_p$
and shifts still farther as $|f|$ increases.  The segment with a
negative slope consists of unstable solutions.

Notice that a sizable portion of the lower branch can also be unstable
because of an intermode scattering~\cite{Carusotto2004} such as shown
in Fig.~\ref{fig:s}(b) for the case of $\bk_p = 0$.  The imaginary
part of the energy $\tilde E(\bk)$ of elementary excitations changes
its sign for some $\bk = \bk_\star$ at a certain threshold point
$|f| = f_\star$, resulting in a spontaneous growth of
$|\psi_{\bk_\star}|$.  Specifically, this occurs when blueshift
$V|\psi_{\bk_p}|^2$ exceeds $\gamma$, whereas at the end of the lower
branch of one-mode solutions the blueshift would have reached a much
greater value of $D/3$ in the limit $\gamma / D \to 0$.  Thus, in the
case of small $\gamma$ one can estimate $\bk_\star$ directly from an
unshifted dispersion law $E(\bk)$ taking into account energy and
momentum conservation [Fig.~\ref{fig:s}(b)].  If $\bk_p = 0$, all
$\bk_\star$ lie on a ring-shaped intersection of the renormalized
energy surface $\tilde E(\bk)$ and pump level
$E = E_p$~\cite{Gavrilov2014.prb.b}.

Since the scattering threshold $f_\star$ is less than the bistability
turning point, the question arises of what exactly happens to the
system when the pump amplitude slightly exceeds~$f_\star$.  Proceeding
from certain analogies in laser physics, one might expect a
second-order phase transition with a continuous amplification of
scattered modes upon increasing $f$, which is indeed quite a common
behavior of dissipative systems in which $\max_\bk \Im \tilde E(\bk)$
smoothly changes its sign in a critical point~\cite{Haken1975.rev}.
However, the answer is different and counter-intuitive: the parametric
break-up of the driven mode is accompanied by a growth rather than
decrease of its own amplitude
$|\psi_{\bk_p}|$~\cite{Gavrilov2014.prb.b}.  This is possible despite
the ``conservative'' kind of the $|\psi|^4$ interaction, because the
system is open.  As a result, the total $|\psi|$ grows spontaneously
even at constant $|f|$ until the blueshift cancels the pump detuning.
Such a process shows a hyperbolic time dependence with a latency
period that tends to infinity for $|f| \to f_\star + 0$ but otherwise
ends up in a singularity point with a very sharp jump of the field.
Analogous scenarios are known as regimes with blowup
(\cite{Landman1988}).

The scattering may lead to an unusual state in which population of
scattered modes $I_s = \sum_{\bk \neq \bk_p} |\psi_\bk|^2$ overcomes
$I_p = |\psi_{\bk_p}|^2$.  When $D / \gamma \gtrsim 10$ and $|f|$ is
close to the threshold, $I_s$ can be several times greater than $I_p$
during a short period just before the jump to the upper branch.  In a
uniform system, the growth of $I_s$ results in the one-mode
instability of the pumped mode (so that its lower-energy state
disappears~\cite{Gavrilov2014.prb.b}), thus, the value of $I_s / I_p$
cannot be abnormally high for a long time.  In the special case of
$\bk_p = 0$, the parametric scattering turns into a purely transient
process that mediates the jump to the upper stability branch and, in
particular, reduces the corresponding threshold at the cost of a
potentially lengthy latency period.

\begin{figure}[b]
  \centering
  \includegraphics[width=\linewidth]{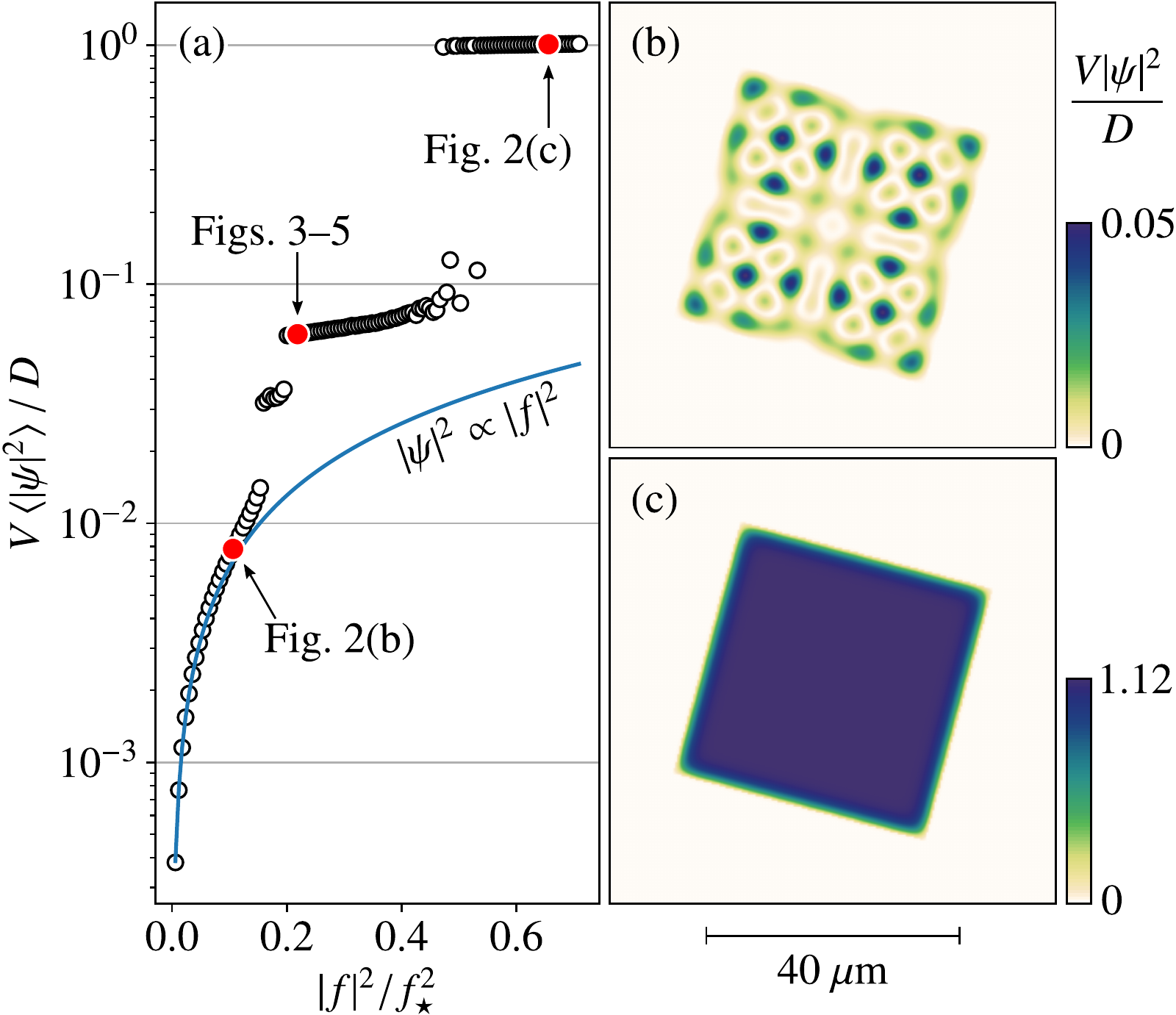}
  \caption{\label{fig:series} (a) Steady-state response to excitation
    in a square microcavity mesa. Parameters are indicated in the main
    text.  Circles represent independent solutions. Intensities
    $|\psi(\br, t)|^2$ are averaged in space (over the mesa) and time
    (over 0.1~ns) after a 1~ns long period of increasing $|f|$ from
    zero to a given amplitude and extra 2 ns long period allotted for
    the establishment of a particular solution at a fixed $|f|$. (b,
    c)~explicit spatial dependences for two solutions.}
\end{figure}

The key idea of this work is that an abnormally high population of
scattered modes (i.\,e., a sort of population inversion for the case
of parametric scattering) can be stabilized owing to a reduced
symmetry.  Let us consider a square quantum well (mesa) with a side of
$L \approx 40~\mu$m and indefinitely high energy barrier at the
boundary.  The main parameters are $\gamma = 0.01$~meV, $D = 0.4$~meV,
and $\bk_p = 0$; the exciton-photon detuning at $\bk = 0$ is zero and,
thus, the polariton mass $m$ is two times larger than the photon mass
$m_\mathrm{ph} = \epsilon E_0 / c^2$, where $E_0 = 1.5$~eV and
$\epsilon = 12.5$ (as in typical GaAs-based microcavities).  We took
into account the non-parabolicity of $E(\bk)$, yet it does not play a
significant role at small $D$.  The interaction constant $V$ only
determines $f_\star \propto V^{-1/2}$ and can be chosen arbitrarily.

Figure~\ref{fig:series}(a) shows a steady-state dependence of an
average $V |\psi|^2$ on $|f|^2$.  Compared to Fig.~\ref{fig:s}(a), it
contains three rather than two stability branches as well as several
isolated points representing transient solutions.  As expected, (i)
the response is linear for $|f| \to 0$ and (ii) the uppermost branch
is characterized by a fully canceled pump detuning
($V |\psi|^2 / D \gtrsim 1$).  This branch is reached at a relatively
small pump power $|f|^2 \approx 0.5 f_\star^2$, which is not
surprising since the presence of sharp potential walls involves the
Rayleigh scattering into different $k$-states with
$\tilde E(\bk) = E_p$; as a result, the polariton density
$\langle |\psi|^2 \rangle$ is higher than in a flat cavity at the same
$f$ unless the system has arrived at the upper branch where all
scattering channels are closed.  It is remarkable, however, that the
response becomes nonlinear already at
$V \langle |\psi|^2 \rangle \lesssim 10^{-2} D$, i.\,e., far below the
threshold value of $V \langle |\psi|^2 \rangle$.  This is explained by
the fact that the field is strongly inhomogeneous even inside the mesa
and has several short-range areas with comparatively high $|\psi|^2$.
As seen in Fig.~\ref{fig:series}(b), the maximum $V |\psi|^2$ equals
$0.05 \, D = 2 \gamma$, which is twice greater than the threshold,
whereas the average $V |\psi|^2$ is still less than $0.008 \, D$.

In turn, the strong inhomogeneity is explained by a reduced rotational
symmetry, owing to which the Rayleigh scattering has certain preferred
directions matching the system geometry.  If a square mesa is oriented
along the $x$ and $y$ axes, the preferred $k$-states are
$(\pm k_R, 0)$ and $(0, \pm k_R)$, where
$k_R \approx \sqrt{2 m D} / \hbar$.  Indeed, the reflection of each of
these waves from a potential wall yields a twin wave with the inverse
$\bk$, whereas all other waves would eventually scatter into many
modes and lose coherence.  The filling of the geometrically preferred
states reveals itself in formation of a semi-periodic standing-wave
pattern whose sharpness appears to be as high as in
Fig.~\ref{fig:series}(b) even at $|f| \to 0$.  Since $k_\star = k_R$
for $k_p = 0$, the already dominant $k$-states (filled via the
Rayleigh scattering) and the corresponding real-space lattice get
amplified parametrically upon increasing $|f|$.  Thus, the onset of
the parametric scattering takes place in a number of small spots
rather than in the whole system at once.

The short-range parametric instability has been studied earlier by
focusing the pump into a 2 $\mu$m spot on the
sample~\cite{Whittaker2017}.  In that case one could no longer
distinguish the pumped and scattered $k$-states and analyze their
interaction in the spirit of Ref.~\cite{Gavrilov2014.prb.b}.
Nonetheless, the bistability effect as well as threshold-like
parametric scattering were observed.  It was found that, in contrast
to the case of one-mode excitation, the onset of the instability does
not end up with a jump to the upper branch in a finite range of pump
powers.  The blowup remains unfinished, because the growth of $|\psi|$
at the spot center makes polaritons more intensively spread out of the
parametrically unstable area, which is equivalent to additional energy
losses.  Such a system permanently remains in a state with many-mode
instability; as a result, it exhibits strong quantum noise and
spontaneous pattern formation~\cite{Whittaker2017}.

The middle branch seen in Fig.~\ref{fig:series}(a) at
$0.2 \lesssim |f|^2 / f_\star^2 \lesssim 0.5$ has a similar nature.
The collective states with strong parametric instability are
stabilized on the average owing to additional losses.  Considering the
real space, one can say that polaritons ballistically move from the
higher-energy ``growth'' areas to the ``decay'' areas.  In the
momentum space, the same process is seen as the scattering into a
continuum of highly dissipative modes.  The dominant $k$-states
suppress the field at the places of their negative interference, which
makes the system inhomogeneous and prevents its continuous transition
to the uppermost branch where the field is, by contrast, spatially
uniform [Fig.~\ref{fig:series}(c)].

In summary, the Rayleigh scattering plays the role of a precursor to
the parametric oscillation transition.  As $f$ is increased, the
system becomes parametrically unstable within a number of spatially
isolated areas where the field is the most strong.  The instability
results in a discontinuous transition accompanied by a sharp growth of
the total intensity.  However, the system remains strongly
inhomogeneous and does not reach the upper branch of solutions in a
wide range of $f$.

\section{Macroscopic loop interaction}
\label{sec:loop}

\begin{figure}
  \centering
  \includegraphics[width=\linewidth]{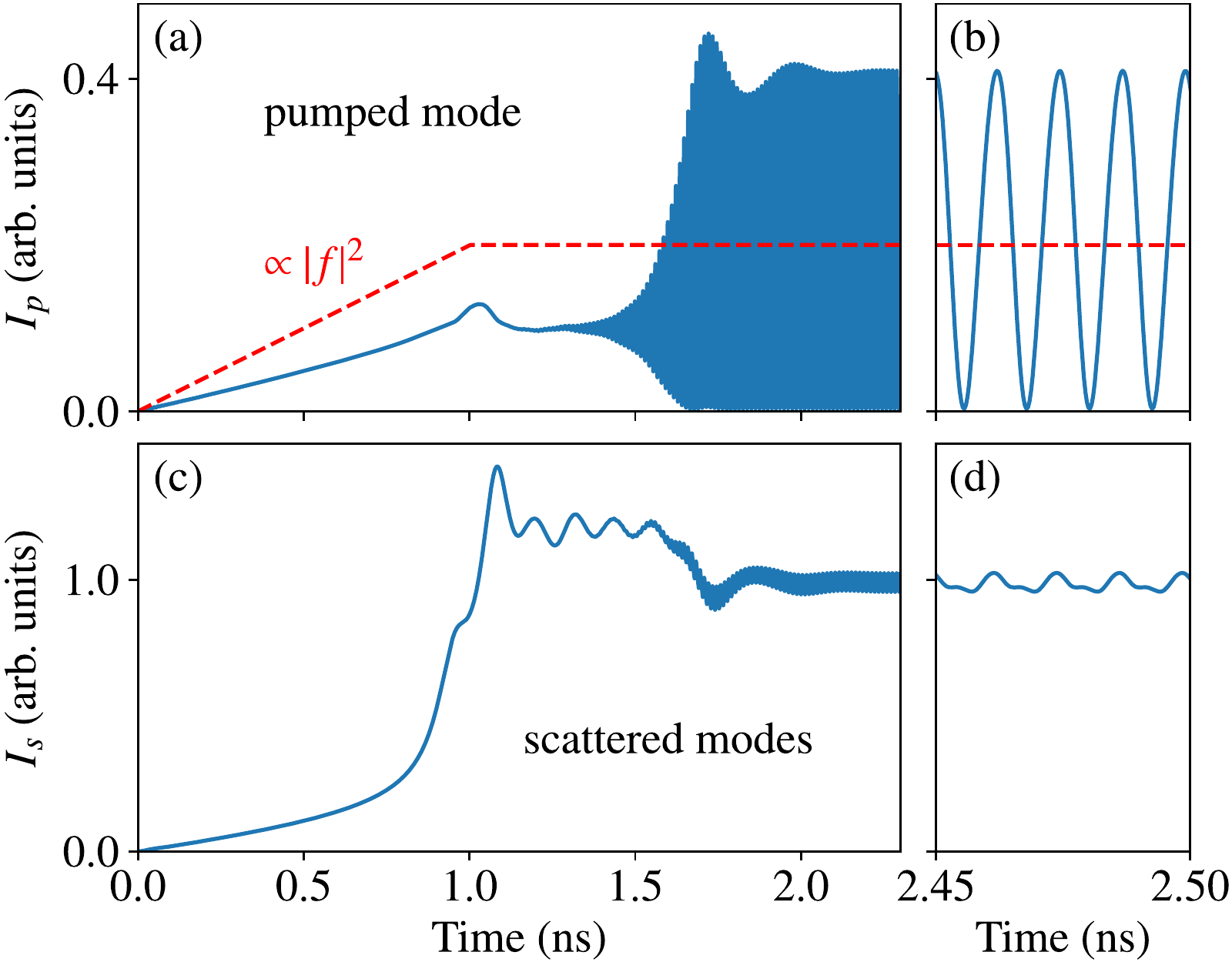}
  \caption{\label{fig:dynamics} Dynamics of the intensities of the
    pumped (a, b) and scattered (c, d) modes.  The pump power $|f|^2$
    is linearly increased in 1~ns from 0 to
    ${\sim} \, 0.22 \, f_\star^2$.  (b) and (d) represent an
    established solution on a more detailed time scale.}
\end{figure}

Let us now turn to a detailed analysis of one characteristic solution
close to the beginning of the middle branch.
Figure~\ref{fig:dynamics} explicitly shows the intensities of the
pumped ($I_p$) and scattered ($I_s$) modes depending on time for
$|f|^2 / f_\star^2 \approx 0.22$.  Technically speaking, $I_p$ is
summed over $k_{x,y} = 0 \pm 0.08~\mu\mathrm{m}^{-1}$ in order to take
account of mode broadening in a confined system and, accordingly,
$I_s$ is summed over the rest of the $k$-space.

The pump is turned on slowly (in 1~ns) to illustrate the onset of the
instability.  Initially, we have $I_p \propto |f|^2$ and
$I_s \propto I_p$ (the Rayleigh scattering is linear).  The increase
of $|f|$ beyond the parametric threshold leads to a significant
increase of $I_s$, which is typical of the first stage of blowup when
much energy is transferred into the system of scattered modes whereas
the pumped mode is increased only slightly~\cite{Gavrilov2014.prb.b}.
As we have argued previously, this process does not involve the entire
area of the mesa, and so the accumulated $I_s$ is still insufficient
for triggering a global one-mode instability of $\bk = 0$.  As a
result, by $t \approx 1.1$~ns the system gets locked in a state with
$I_s / I_p \gtrsim 10$.

Since the normal way out of the many-mode instability is unfeasible,
qualitatively new collective phenomena come into being.  In
particular, the pumped mode exhibits self-pulsations whose amplitude
rapidly grows in the interval from $t \approx 1.3$~ns to $1.7$~ns,
after which a fairly regular oscillation regime is established in
several hundreds of picoseconds.  This effect is naturally explained
by the filling of a new coherent mode with the same $\bk = \bk_p = 0$
but different frequency.  The oscillation period $T \approx 12$~ps
approximately matches the inverse pump detuning $h / D \approx 10$~ps,
which means that the new mode is located slightly above the ground
polariton state.

\begin{figure}
  \centering
  \includegraphics[width=\linewidth]{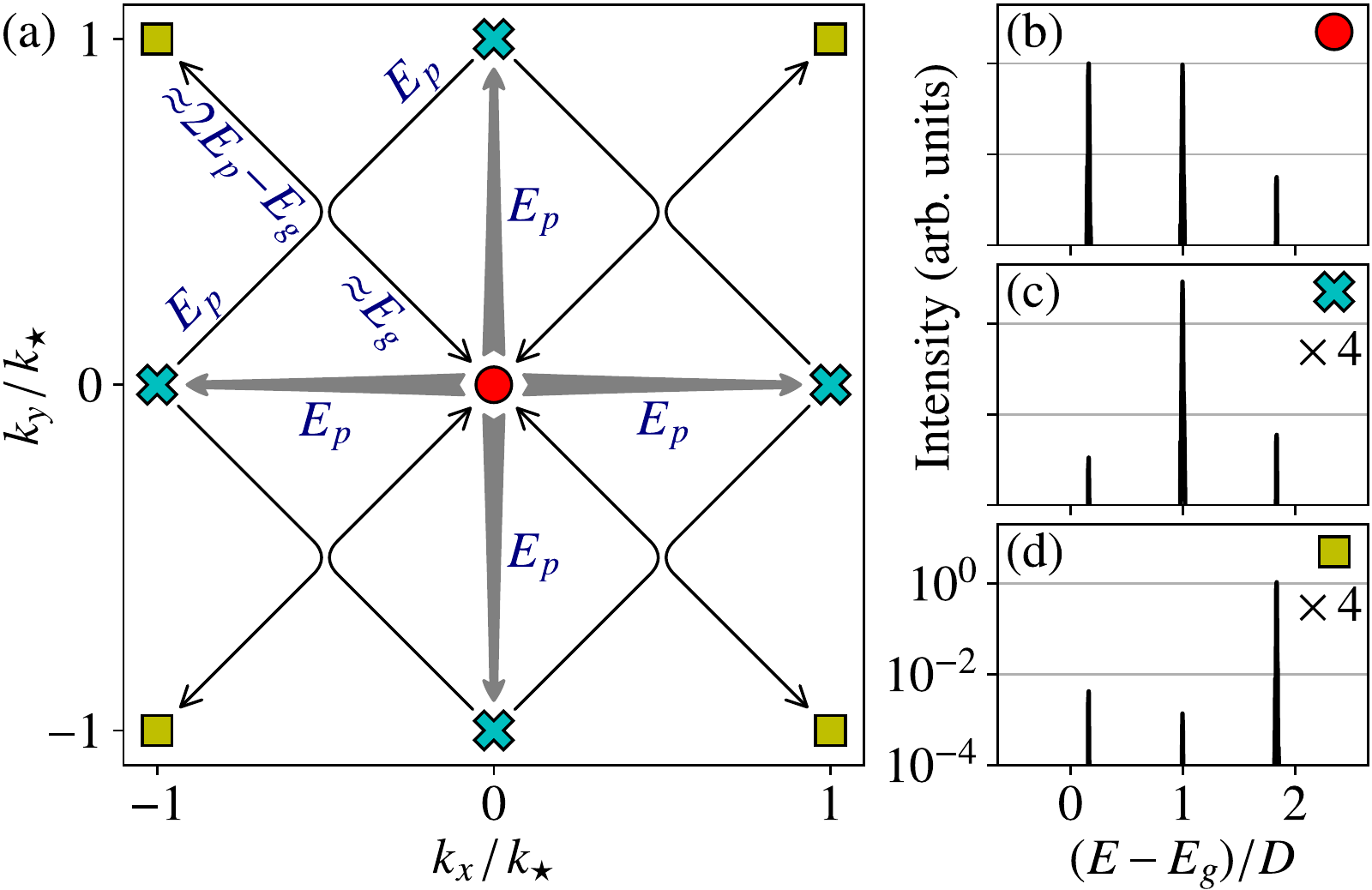}
  \caption{\label{fig:spectra} (a) Scheme of the loop parametric
    scattering.  (b--d) The spectra of the main $k$-states at the
    stage of regular oscillations.}
\end{figure}

The emergence of a new coherent polariton state with $\bk = \bk_p$ and
$E < E_p$ is a very uncommon phenomenon.  Being somewhat analogous to
dynamical condensation~\cite{Sun2012}, it is hardly expected under
coherent driving, because such states cannot be excited via scattering
from the pumped mode~\cite{Gavrilov2020.usp.en}.  If $\bk_p = 0$, the
direct scattering leads only to the states with $\tilde E(\bk) = E_p$,
whereas all other two-particle processes are usually weak and do not
reach the threshold of the parametric amplification.  However, the
abnormal population of scattered modes makes some of the indirect
interaction channels particularly strong.

The diagram in Fig.~\ref{fig:spectra}(a) represents the interaction
process resulting in the filling of the $k = 0$ mode near $E = E_g$.
Specifically, each adjacent pair of the geometrically preferred
states, e.\,g., $\bk_1 = (\pm k_\star, 0)$ and
$\bk_2 = (0, \pm k_\star)$, scatters into $(0, 0)$ and
$\bk_1 + \bk_2 = (\pm k_\star, \pm k_\star)$, which is consistent with
energy conservation in the vicinity of the ground state where the
polariton dispersion law is nearly parabolic.  The overall scheme
comprises a number of two-particle interactions, however, each of the
$k_\star$ states acts as a source in two scattering processes at once,
whereas the $k = 0$ state occurs to be the joint target of all of
them, which necessarily implies phase synchronization of different
interaction channels.

Figures~\ref{fig:spectra}(b)--(d) show the spectra of the main
$k$-states engaged in the loop scattering, which are obtained by the
Fourier transform of $\psi_\bk(t)$ over 2~ns at the stage of regular
oscillations.  As expected, the $k = 0$ state
[Fig.~\ref{fig:spectra}(b)] has two peaks at $E = E_p$ and
$E \gtrsim E_g$ whose intensities are nearly the same, in agreement
with the fact that $I_p(t)$ drops down to zero at the oscillation
minima.  The geometrically preferred modes $k = k_\star$
[Fig.~\ref{fig:spectra}(c)] have, by contrast, only one strong peak at
$E = E_p$ and thus appear to be particularly steady.  The ``idlers''
with $k = \sqrt{2} k_\star$ [Fig.~\ref{fig:spectra}(d)] show the peak
at $E \lesssim 2 E_p - E_g$ whose total intensity (summed over 4
modes) nearly equals the intensity of the driven mode at $E = E_p$.
Notice that all spectral lines are almost unbroadened, which is
indicative of their ``parametric'' nature and precise synchronization
of the respective $k$-states.  The spectra also contain sharp peaks
near $(E - E_g) / D = 3,4,$ etc.\ (not shown), however, that peaks are
located far from resonances and have thus comparatively low
amplitudes.

\begin{figure}
  \centering
  \includegraphics[width=\linewidth]{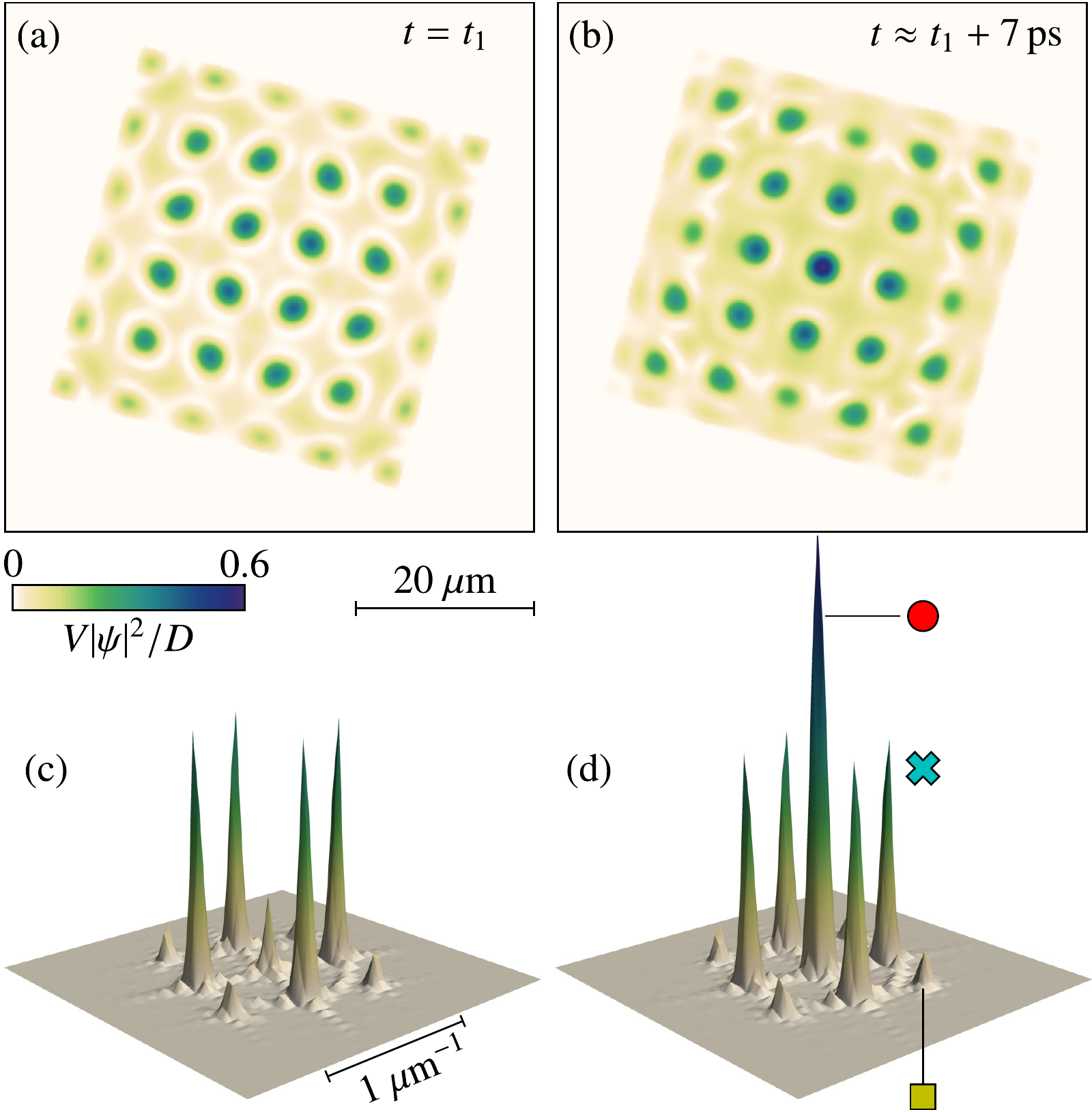}
  \caption{\label{fig:space} Typical real-space (a, b) and
    momentum-space (c, d) field patterns at the stage of regular
    oscillations for two time instants separated by 7~ps
    (approximately half of a period).  The full evolution can be seen
    in the Supplemental Material~\cite{Note1}.}
\end{figure}

The explicit real- and momentum-space distributions of $|\psi|^2$ are
shown in Fig.~\ref{fig:space} for two time instants which are nearly
half a period apart; the corresponding continuous evolution is
displayed by a separate video file~~\cite{Note1}.  Is seen that all
modes in the $k$-space except $k = 0$ remain nearly steady.  Some of
them are populated significantly, thus, each change in $\psi_0$ has to
be accompanied by a redistribution of $|\psi(\br)|$ in the real space
in view of interference.  As seen from comparison of
Figs.~\ref{fig:space}(a) and (b), the growth of $I_p = |\psi_0|^2$
leads to the shift of the spatial lattice by nearly half of a period
in both directions, so that all maxima approximately turn into minima
and vice versa. Notice as well that the maximum of
$V|\psi(\br, t)|^2$, which is attained at the center of
Fig.~\ref{fig:space}(b), is no greater than $0.6 \, D$ and, therefore,
the upper-branch states are not yet feasible even at a single point.

It is remarkable that rapid changes of $I_p$ at the regular stage
hardly affect $I_s$, which might seem untypical for a system with
strong parametric instability of the pumped mode.  The observed
steadiness of $I_s$ results in a constant rate of back scattering and,
eventually, in a fully regular character of spatiotemporal
oscillations.  In its turn, it stems from the abnormal population:
since the pumped mode gives a comparatively small average contribution
to the blueshift, $I_s$ varies on the scale of the polariton lifetime
$\tau = \hbar / \gamma$ that largely exceeds the oscillation period
$T \gtrsim h / D$ as long as $\gamma \ll D$.  The increase in
$\langle I_p \rangle / \langle I_s \rangle$ would lead to an increased
oscillation amplitude of $I_s$ and, thus, increased volatility of the
back scattering.  As a result, the system experiences a transition to
dynamical chaos.  For the series of solutions represented in
Fig.~\ref{fig:series}(a), this occurs in the last quarter of the
middle branch where ratio $\langle I_p \rangle / \langle I_s \rangle$
becomes nearly two times greater compared to the beginning of the same
branch.  Analysis of chaotic solutions is beyond the scope of this
paper; we only notice that chaos makes the transition to the upper
branch partially accidental, so that the two branches overlap within a
finite range of pump powers around $0.5 \, f_\star^2$.

Turning back to self-pulsations, notice that the two main energy peaks
of the $k = 0$ mode have equal intensities for most of the middle
branch.  Specifically, this is true in the range of $f^2$ from
$0.2 f_\star^2$ (where that branch begins) to $0.4 f_\star^2$ (where
the system experiences the first period-doubling bifurcation on its
way to chaos).  The equality of energy peaks and the corresponding
full-scale oscillation of $I_p$ are indicative of a precise balance of
the direct and inverse scattering processes.

The length of the middle branch depends on the system size.  The
necessary condition of the considered phenomena is a strong
inhomogeneity achieved already at $|f| \to 0$ owing to formation of
standing waves.  In other words, the geometrically preferred
$k$-states must be significantly stronger than all other modes with
$k \neq 0$.  This condition is met when the system size $L$ is
comparable to the distance traveled by the $k_\star$ polaritons on the
scale of their lifetime.  When that is the case, all $k$-states become
highly dissipative through multiple reflections from the potential
walls, except for the ``preferred'' states which run perpendicular to
one of them and form standing waves (generally speaking, this is true
for all regular polygons of a reasonably small order).  Thus, an
increase of the $Q$-factor allows one to extend the range of the
system sizes suitable to achieve the abnormal population and
macroscopic loop scattering.  The increase of $Q$ also conforms to the
other assumption that $\gamma \ll D$.  It is worth noting that a state
in which blueshift $V |\psi|^2 \sim D$ exceeds $\gamma$ by more than
10 times with no side effects such as nonlinear losses seems to be
specific for polaritonic systems with strong exciton-photon coupling.

\section{Discussion}
\label{sec:discussion}

Up to date, the parametric scattering of cavity polaritons is often
thought to be just a macroscopic effect of the two-particle
interaction with well-defined ``source'' and ``target'' $k$-states.
It was found, however, that for a greater $D / \gamma$ the parametric
scattering is an essentially collective process that successively
involves many modes with different wave numbers $|\bk|$ even when the
pump amplitude is arbitrarily close to the
threshold~\cite{Gavrilov2014.prb.b,Gavrilov2015,Gavrilov2020.usp.en}.
The most unstable many-mode states exhibit the abnormal population of
scattered modes.  As a rule, they are transient and only mediate the
jump to the upper stability branch (for $k_p = 0$) or to a
well-developed OPO regime with strong signal and idler modes.  The aim
of this work was to find a way to make the abnormal population
persistent under constant driving conditions in a spatially extended
polariton system.  We have found it possible in the presence of a
reduced rotational symmetry.  The asymmetric (e.\,g., polygonal)
systems show the macroscopic loop interaction in which several
$k$-states simultaneously act as the ``sources'' and ``targets'' of
the parametric scattering.

The considered phase transition is somewhat analogous to the dynamical
condensation~\cite{Sun2012}.  Indeed, when the $k \neq 0$ modes, which
are neither pumped nor energetically favored, acquire an abnormally
high intensity, their multiple interactions result in the appearance
of a new coherent mode near the ground-state level.  The analogy with
BECs or lasers is imperfect, however, because the new condensate is
populated parametrically---in contrast to~\cite{Sun2012}, where its
formation has statistical reasons.  At the same time, we have found
the ``condensation'' to be accompanied by an induced synchronization
of several back scattering channels, which is a relatively complex
process that lies beyond the scope of plain two-particle
interactions. The greater the polygon order, the greater the number of
highly populated modes that become synchronized.  Calculations show
that even a purely circular shape of the microcavity does not prevent
massive back scattering which in this case proceeds through the
spontaneous breakdown of the ring symmetry and results in chaotic
dynamics.  The statistical and, possibly, quantum aspects of the
synchronized loop scattering have yet to be studied.

The self-pulsation of a polariton fluid under coherent driving is also
a remarkable process that is usually prevented by the dissipative
nature of polaritons combined with a purely repulsive kind of their
interaction (as opposed to lasers~\cite{Cross1993,Staliunas.book} and
systems with self-focusing~\cite{Ferre2017}).  Even when the ground
state is split into two spin or Josephson sublevels, both components
of the condensate have the same ``forced'' energy $E_p$ and thus do
not oscillate at $|f| \to 0$ as well as $|f| \to \infty$.  However,
the combination of the ground-state splitting and nonlinearity does
result in regular or chaotic oscillations in certain particular
cases~\cite{Sarchi2008, Solnyshkov2009.j, Gavrilov2016, Leblanc2020}.
When $V$ is nonzero and the spin splitting significantly exceeds
$\gamma$, all one-mode states are forbidden in a finite range of
$|f|$, resulting in chaos, dipolar networks, chimera states, and
spontaneously formed vortices even for a purely uniform polariton
system pumped by a plane wave~\cite{Gavrilov2016, Gavrilov2018,
  Gavrilov2020}.  These phenomena are underlaid by a different kind of
loop scattering that takes place in the presence of linear coupling of
opposite spins~\cite{Gavrilov2017.en,Gavrilov2020.usp.en}.  By
contrast, the system considered in the current work is effectively
scalar, implying that all polaritons have just the same spin, which
corresponds to the case of circularly polarized
excitation~\cite{Shelykh2010, Vladimirova2010, Gavrilov2010.jetpl.en,
  Sekretenko2013.10ps}.  In contrast to another recent study, in which
the second mode comes into play owing to size quantization in a
$\mu$m-sized micropillar~\cite{Leblanc2020}, our system has a large
spatial extent and the respective mode splitting is fairly negligible.
Thus, the second condensate appears at $k = 0$ purely dynamically.  At
last, we notice that regular oscillations represent only the simplest
kind of evolution, whereas the increase of $|f|$ or more complex shape
of a microcavity result in new collective phenomena that call for
investigation.

\begin{acknowledgments}
  I am grateful to V.\,D.~Kulakovskii for stimulating discussions.
  The work was supported by the Russian Foundation for Basic Research
  (Grant No.\ 19-02-00988) and Volkswagen Foundation (Grant No.\
  97758).
\end{acknowledgments}

%


\begin{thebibliography}{46}%
\makeatletter
\providecommand \@ifxundefined [1]{%
 \@ifx{#1\undefined}
}%
\providecommand \@ifnum [1]{%
 \ifnum #1\expandafter \@firstoftwo
 \else \expandafter \@secondoftwo
 \fi
}%
\providecommand \@ifx [1]{%
 \ifx #1\expandafter \@firstoftwo
 \else \expandafter \@secondoftwo
 \fi
}%
\providecommand \natexlab [1]{#1}%
\providecommand \enquote  [1]{``#1''}%
\providecommand \bibnamefont  [1]{#1}%
\providecommand \bibfnamefont [1]{#1}%
\providecommand \citenamefont [1]{#1}%
\providecommand \href@noop [0]{\@secondoftwo}%
\providecommand \href [0]{\begingroup \@sanitize@url \@href}%
\providecommand \@href[1]{\@@startlink{#1}\@@href}%
\providecommand \@@href[1]{\endgroup#1\@@endlink}%
\providecommand \@sanitize@url [0]{\catcode `\\12\catcode `\$12\catcode
  `\&12\catcode `\#12\catcode `\^12\catcode `\_12\catcode `\%12\relax}%
\providecommand \@@startlink[1]{}%
\providecommand \@@endlink[0]{}%
\providecommand \url  [0]{\begingroup\@sanitize@url \@url }%
\providecommand \@url [1]{\endgroup\@href {#1}{\urlprefix }}%
\providecommand \urlprefix  [0]{URL }%
\providecommand \Eprint [0]{\href }%
\providecommand \doibase [0]{https://doi.org/}%
\providecommand \selectlanguage [0]{\@gobble}%
\providecommand \bibinfo  [0]{\@secondoftwo}%
\providecommand \bibfield  [0]{\@secondoftwo}%
\providecommand \translation [1]{[#1]}%
\providecommand \BibitemOpen [0]{}%
\providecommand \bibitemStop [0]{}%
\providecommand \bibitemNoStop [0]{.\EOS\space}%
\providecommand \EOS [0]{\spacefactor3000\relax}%
\providecommand \BibitemShut  [1]{\csname bibitem#1\endcsname}%
\let\auto@bib@innerbib\@empty
\bibitem [{\citenamefont {Weisbuch}\ \emph {et~al.}(1992)\citenamefont
  {Weisbuch}, \citenamefont {Nishioka}, \citenamefont {Ishikawa},\ and\
  \citenamefont {Arakawa}}]{Weisbuch1992}%
  \BibitemOpen
  \bibfield  {author} {\bibinfo {author} {\bibfnamefont {C.}~\bibnamefont
  {Weisbuch}}, \bibinfo {author} {\bibfnamefont {M.}~\bibnamefont {Nishioka}},
  \bibinfo {author} {\bibfnamefont {A.}~\bibnamefont {Ishikawa}},\ and\
  \bibinfo {author} {\bibfnamefont {Y.}~\bibnamefont {Arakawa}},\ }\href
  {https://doi.org/10.1103/PhysRevLett.69.3314} {\bibfield  {journal} {\bibinfo
   {journal} {Phys. Rev. Lett.}\ }\textbf {\bibinfo {volume} {69}},\ \bibinfo
  {pages} {3314} (\bibinfo {year} {1992})}\BibitemShut {NoStop}%
\bibitem [{\citenamefont {Yamamoto}\ \emph {et~al.}(2000)\citenamefont
  {Yamamoto}, \citenamefont {Tassone},\ and\ \citenamefont
  {Cao}}]{Yamamoto.book}%
  \BibitemOpen
  \bibfield  {author} {\bibinfo {author} {\bibfnamefont {Y.}~\bibnamefont
  {Yamamoto}}, \bibinfo {author} {\bibfnamefont {T.}~\bibnamefont {Tassone}},\
  and\ \bibinfo {author} {\bibfnamefont {H.}~\bibnamefont {Cao}},\ }\href@noop
  {} {\emph {\bibinfo {title} {Semiconductor Cavity Quantum Electrodynamics}}}\
  (\bibinfo  {publisher} {Springer},\ \bibinfo {address} {Berlin},\ \bibinfo
  {year} {2000})\BibitemShut {NoStop}%
\bibitem [{\citenamefont {Kavokin}\ \emph {et~al.}(2017)\citenamefont
  {Kavokin}, \citenamefont {Baumberg}, \citenamefont {Malpuech},\ and\
  \citenamefont {Laussy}}]{Kavokin.book.2017}%
  \BibitemOpen
  \bibfield  {author} {\bibinfo {author} {\bibfnamefont {A.~V.}\ \bibnamefont
  {Kavokin}}, \bibinfo {author} {\bibfnamefont {J.~J.}\ \bibnamefont
  {Baumberg}}, \bibinfo {author} {\bibfnamefont {G.}~\bibnamefont {Malpuech}},\
  and\ \bibinfo {author} {\bibfnamefont {P.}~\bibnamefont {Laussy}},\
  }\href@noop {} {\emph {\bibinfo {title} {Microcavities}}},\ \bibinfo
  {edition} {2nd}\ ed.\ (\bibinfo  {publisher} {Oxford University Press},\
  \bibinfo {address} {New York},\ \bibinfo {year} {2017})\BibitemShut {NoStop}%
\bibitem [{\citenamefont {Kasprzak}\ \emph {et~al.}(2006)\citenamefont
  {Kasprzak}, \citenamefont {Richard}, \citenamefont {Kundermann},
  \citenamefont {Baas}, \citenamefont {Jeambrun}, \citenamefont {Keeling},
  \citenamefont {Marchetti}, \citenamefont {Szyma{\'n}ska}, \citenamefont
  {Andr{\'e}}, \citenamefont {Staehli}, \citenamefont {Savona}, \citenamefont
  {Littlewood}, \citenamefont {Deveaud},\ and\ \citenamefont
  {Dang}}]{Kasprzak2006}%
  \BibitemOpen
  \bibfield  {author} {\bibinfo {author} {\bibfnamefont {J.}~\bibnamefont
  {Kasprzak}}, \bibinfo {author} {\bibfnamefont {M.}~\bibnamefont {Richard}},
  \bibinfo {author} {\bibfnamefont {S.}~\bibnamefont {Kundermann}}, \bibinfo
  {author} {\bibfnamefont {A.}~\bibnamefont {Baas}}, \bibinfo {author}
  {\bibfnamefont {P.}~\bibnamefont {Jeambrun}}, \bibinfo {author}
  {\bibfnamefont {J.~M.~J.}\ \bibnamefont {Keeling}}, \bibinfo {author}
  {\bibfnamefont {F.~M.}\ \bibnamefont {Marchetti}}, \bibinfo {author}
  {\bibfnamefont {M.~H.}\ \bibnamefont {Szyma{\'n}ska}}, \bibinfo {author}
  {\bibfnamefont {R.}~\bibnamefont {Andr{\'e}}}, \bibinfo {author}
  {\bibfnamefont {J.~L.}\ \bibnamefont {Staehli}}, \bibinfo {author}
  {\bibfnamefont {V.}~\bibnamefont {Savona}}, \bibinfo {author} {\bibfnamefont
  {P.~B.}\ \bibnamefont {Littlewood}}, \bibinfo {author} {\bibfnamefont
  {B.}~\bibnamefont {Deveaud}},\ and\ \bibinfo {author} {\bibfnamefont {L.~S.}\
  \bibnamefont {Dang}},\ }\href {https://doi.org/10.1038/nature05131}
  {\bibfield  {journal} {\bibinfo  {journal} {Nature}\ }\textbf {\bibinfo
  {volume} {443}},\ \bibinfo {pages} {409} (\bibinfo {year}
  {2006})}\BibitemShut {NoStop}%
\bibitem [{\citenamefont {Baas}\ \emph {et~al.}(2006)\citenamefont {Baas},
  \citenamefont {Karr}, \citenamefont {Romanelli}, \citenamefont {Bramati},\
  and\ \citenamefont {Giacobino}}]{Baas2006}%
  \BibitemOpen
  \bibfield  {author} {\bibinfo {author} {\bibfnamefont {A.}~\bibnamefont
  {Baas}}, \bibinfo {author} {\bibfnamefont {J.-P.}\ \bibnamefont {Karr}},
  \bibinfo {author} {\bibfnamefont {M.}~\bibnamefont {Romanelli}}, \bibinfo
  {author} {\bibfnamefont {A.}~\bibnamefont {Bramati}},\ and\ \bibinfo {author}
  {\bibfnamefont {E.}~\bibnamefont {Giacobino}},\ }\href
  {https://doi.org/10.1103/PhysRevLett.96.176401} {\bibfield  {journal}
  {\bibinfo  {journal} {Phys. Rev. Lett.}\ }\textbf {\bibinfo {volume} {96}},\
  \bibinfo {pages} {176401} (\bibinfo {year} {2006})}\BibitemShut {NoStop}%
\bibitem [{\citenamefont {Ciuti}\ \emph {et~al.}(2003)\citenamefont {Ciuti},
  \citenamefont {Schwendimann},\ and\ \citenamefont {Quattropani}}]{Ciuti2003}%
  \BibitemOpen
  \bibfield  {author} {\bibinfo {author} {\bibfnamefont {C.}~\bibnamefont
  {Ciuti}}, \bibinfo {author} {\bibfnamefont {P.}~\bibnamefont
  {Schwendimann}},\ and\ \bibinfo {author} {\bibfnamefont {A.}~\bibnamefont
  {Quattropani}},\ }\href {http://stacks.iop.org/ss/18/S279} {\bibfield
  {journal} {\bibinfo  {journal} {Semicond. Sci. Technol.}\ }\textbf {\bibinfo
  {volume} {18}},\ \bibinfo {pages} {S279} (\bibinfo {year}
  {2003})}\BibitemShut {NoStop}%
\bibitem [{\citenamefont {Gippius}\ \emph {et~al.}(2004)\citenamefont
  {Gippius}, \citenamefont {Tikhodeev}, \citenamefont {Kulakovskii},
  \citenamefont {Krizhanovskii},\ and\ \citenamefont
  {Tartakovskii}}]{Gippius2004.epl}%
  \BibitemOpen
  \bibfield  {author} {\bibinfo {author} {\bibfnamefont {N.~A.}\ \bibnamefont
  {Gippius}}, \bibinfo {author} {\bibfnamefont {S.~G.}\ \bibnamefont
  {Tikhodeev}}, \bibinfo {author} {\bibfnamefont {V.~D.}\ \bibnamefont
  {Kulakovskii}}, \bibinfo {author} {\bibfnamefont {D.~N.}\ \bibnamefont
  {Krizhanovskii}},\ and\ \bibinfo {author} {\bibfnamefont {A.~I.}\
  \bibnamefont {Tartakovskii}},\ }\href
  {http://stacks.iop.org/0295-5075/67/i=6/a=997} {\bibfield  {journal}
  {\bibinfo  {journal} {Europhys. Lett.}\ }\textbf {\bibinfo {volume} {67}},\
  \bibinfo {pages} {997} (\bibinfo {year} {2004})}\BibitemShut {NoStop}%
\bibitem [{\citenamefont {Carusotto}\ and\ \citenamefont
  {Ciuti}(2013)}]{Carusotto2013}%
  \BibitemOpen
  \bibfield  {author} {\bibinfo {author} {\bibfnamefont {I.}~\bibnamefont
  {Carusotto}}\ and\ \bibinfo {author} {\bibfnamefont {C.}~\bibnamefont
  {Ciuti}},\ }\href {https://doi.org/10.1103/RevModPhys.85.299} {\bibfield
  {journal} {\bibinfo  {journal} {Rev. Mod. Phys.}\ }\textbf {\bibinfo {volume}
  {85}},\ \bibinfo {pages} {299} (\bibinfo {year} {2013})}\BibitemShut
  {NoStop}%
\bibitem [{\citenamefont {Savvidis}\ \emph {et~al.}(2000)\citenamefont
  {Savvidis}, \citenamefont {Baumberg}, \citenamefont {Stevenson},
  \citenamefont {Skolnick}, \citenamefont {Whittaker},\ and\ \citenamefont
  {Roberts}}]{Savvidis2000}%
  \BibitemOpen
  \bibfield  {author} {\bibinfo {author} {\bibfnamefont {P.~G.}\ \bibnamefont
  {Savvidis}}, \bibinfo {author} {\bibfnamefont {J.~J.}\ \bibnamefont
  {Baumberg}}, \bibinfo {author} {\bibfnamefont {R.~M.}\ \bibnamefont
  {Stevenson}}, \bibinfo {author} {\bibfnamefont {M.~S.}\ \bibnamefont
  {Skolnick}}, \bibinfo {author} {\bibfnamefont {D.~M.}\ \bibnamefont
  {Whittaker}},\ and\ \bibinfo {author} {\bibfnamefont {J.~S.}\ \bibnamefont
  {Roberts}},\ }\href {https://doi.org/10.1103/PhysRevLett.84.1547} {\bibfield
  {journal} {\bibinfo  {journal} {Phys. Rev. Lett.}\ }\textbf {\bibinfo
  {volume} {84}},\ \bibinfo {pages} {1547} (\bibinfo {year}
  {2000})}\BibitemShut {NoStop}%
\bibitem [{\citenamefont {Stevenson}\ \emph {et~al.}(2000)\citenamefont
  {Stevenson}, \citenamefont {Astratov}, \citenamefont {Skolnick},
  \citenamefont {Whittaker}, \citenamefont {Emam-Ismail}, \citenamefont
  {Tartakovskii}, \citenamefont {Savvidis}, \citenamefont {Baumberg},\ and\
  \citenamefont {Roberts}}]{Stevenson2000}%
  \BibitemOpen
  \bibfield  {author} {\bibinfo {author} {\bibfnamefont {R.~M.}\ \bibnamefont
  {Stevenson}}, \bibinfo {author} {\bibfnamefont {V.~N.}\ \bibnamefont
  {Astratov}}, \bibinfo {author} {\bibfnamefont {M.~S.}\ \bibnamefont
  {Skolnick}}, \bibinfo {author} {\bibfnamefont {D.~M.}\ \bibnamefont
  {Whittaker}}, \bibinfo {author} {\bibfnamefont {M.}~\bibnamefont
  {Emam-Ismail}}, \bibinfo {author} {\bibfnamefont {A.~I.}\ \bibnamefont
  {Tartakovskii}}, \bibinfo {author} {\bibfnamefont {P.~G.}\ \bibnamefont
  {Savvidis}}, \bibinfo {author} {\bibfnamefont {J.~J.}\ \bibnamefont
  {Baumberg}},\ and\ \bibinfo {author} {\bibfnamefont {J.~S.}\ \bibnamefont
  {Roberts}},\ }\href {https://doi.org/10.1103/PhysRevLett.85.3680} {\bibfield
  {journal} {\bibinfo  {journal} {Phys. Rev. Lett.}\ }\textbf {\bibinfo
  {volume} {85}},\ \bibinfo {pages} {3680} (\bibinfo {year}
  {2000})}\BibitemShut {NoStop}%
\bibitem [{\citenamefont {Butt{\'e}}\ \emph {et~al.}(2003)\citenamefont
  {Butt{\'e}}, \citenamefont {Skolnick}, \citenamefont {Whittaker},
  \citenamefont {Bajoni},\ and\ \citenamefont {Roberts}}]{Butte2003}%
  \BibitemOpen
  \bibfield  {author} {\bibinfo {author} {\bibfnamefont {R.}~\bibnamefont
  {Butt{\'e}}}, \bibinfo {author} {\bibfnamefont {M.~S.}\ \bibnamefont
  {Skolnick}}, \bibinfo {author} {\bibfnamefont {D.~M.}\ \bibnamefont
  {Whittaker}}, \bibinfo {author} {\bibfnamefont {D.}~\bibnamefont {Bajoni}},\
  and\ \bibinfo {author} {\bibfnamefont {J.~S.}\ \bibnamefont {Roberts}},\
  }\href {https://doi.org/10.1103/PhysRevB.68.115325} {\bibfield  {journal}
  {\bibinfo  {journal} {Phys. Rev. B}\ }\textbf {\bibinfo {volume} {68}},\
  \bibinfo {pages} {115325} (\bibinfo {year} {2003})}\BibitemShut {NoStop}%
\bibitem [{\citenamefont {Whittaker}(2001)}]{Whittaker2001}%
  \BibitemOpen
  \bibfield  {author} {\bibinfo {author} {\bibfnamefont {D.~M.}\ \bibnamefont
  {Whittaker}},\ }\href {https://doi.org/10.1103/PhysRevB.63.193305} {\bibfield
   {journal} {\bibinfo  {journal} {Phys. Rev. B}\ }\textbf {\bibinfo {volume}
  {63}},\ \bibinfo {pages} {193305} (\bibinfo {year} {2001})}\BibitemShut
  {NoStop}%
\bibitem [{\citenamefont {Ciuti}\ \emph {et~al.}(2001)\citenamefont {Ciuti},
  \citenamefont {Schwendimann},\ and\ \citenamefont {Quattropani}}]{Ciuti2001}%
  \BibitemOpen
  \bibfield  {author} {\bibinfo {author} {\bibfnamefont {C.}~\bibnamefont
  {Ciuti}}, \bibinfo {author} {\bibfnamefont {P.}~\bibnamefont
  {Schwendimann}},\ and\ \bibinfo {author} {\bibfnamefont {A.}~\bibnamefont
  {Quattropani}},\ }\href {https://doi.org/10.1103/PhysRevB.63.041303}
  {\bibfield  {journal} {\bibinfo  {journal} {Phys. Rev. B}\ }\textbf {\bibinfo
  {volume} {63}},\ \bibinfo {pages} {041303} (\bibinfo {year}
  {2001})}\BibitemShut {NoStop}%
\bibitem [{\citenamefont {Dagvadorj}\ \emph {et~al.}(2015)\citenamefont
  {Dagvadorj}, \citenamefont {Fellows}, \citenamefont {{Matyja\ifmmode
  \acute{s}\else {\'s}\fi{}kiewicz}}, \citenamefont {Marchetti}, \citenamefont
  {Carusotto},\ and\ \citenamefont {{Szyma\ifmmode \acute{n}\else
  {\'n}\fi{}ska}}}]{Dagvadorj2015}%
  \BibitemOpen
  \bibfield  {author} {\bibinfo {author} {\bibfnamefont {G.}~\bibnamefont
  {Dagvadorj}}, \bibinfo {author} {\bibfnamefont {J.~M.}\ \bibnamefont
  {Fellows}}, \bibinfo {author} {\bibfnamefont {S.}~\bibnamefont
  {{Matyja\ifmmode \acute{s}\else {\'s}\fi{}kiewicz}}}, \bibinfo {author}
  {\bibfnamefont {F.~M.}\ \bibnamefont {Marchetti}}, \bibinfo {author}
  {\bibfnamefont {I.}~\bibnamefont {Carusotto}},\ and\ \bibinfo {author}
  {\bibfnamefont {M.~H.}\ \bibnamefont {{Szyma\ifmmode \acute{n}\else
  {\'n}\fi{}ska}}},\ }\href {https://doi.org/10.1103/PhysRevX.5.041028}
  {\bibfield  {journal} {\bibinfo  {journal} {Phys. Rev. X}\ }\textbf {\bibinfo
  {volume} {5}},\ \bibinfo {pages} {041028} (\bibinfo {year}
  {2015})}\BibitemShut {NoStop}%
\bibitem [{\citenamefont {Savvidis}\ \emph {et~al.}(2001)\citenamefont
  {Savvidis}, \citenamefont {Ciuti}, \citenamefont {Baumberg}, \citenamefont
  {Whittaker}, \citenamefont {Skolnick},\ and\ \citenamefont
  {Roberts}}]{Savvidis2001}%
  \BibitemOpen
  \bibfield  {author} {\bibinfo {author} {\bibfnamefont {P.~G.}\ \bibnamefont
  {Savvidis}}, \bibinfo {author} {\bibfnamefont {C.}~\bibnamefont {Ciuti}},
  \bibinfo {author} {\bibfnamefont {J.~J.}\ \bibnamefont {Baumberg}}, \bibinfo
  {author} {\bibfnamefont {D.~M.}\ \bibnamefont {Whittaker}}, \bibinfo {author}
  {\bibfnamefont {M.~S.}\ \bibnamefont {Skolnick}},\ and\ \bibinfo {author}
  {\bibfnamefont {J.~S.}\ \bibnamefont {Roberts}},\ }\href
  {https://doi.org/10.1103/PhysRevB.64.075311} {\bibfield  {journal} {\bibinfo
  {journal} {Phys. Rev. B}\ }\textbf {\bibinfo {volume} {64}},\ \bibinfo
  {pages} {075311} (\bibinfo {year} {2001})}\BibitemShut {NoStop}%
\bibitem [{\citenamefont {Whittaker}(2005)}]{Whittaker2005}%
  \BibitemOpen
  \bibfield  {author} {\bibinfo {author} {\bibfnamefont {D.~M.}\ \bibnamefont
  {Whittaker}},\ }\href {https://doi.org/10.1103/PhysRevB.71.115301} {\bibfield
   {journal} {\bibinfo  {journal} {Phys. Rev. B}\ }\textbf {\bibinfo {volume}
  {71}},\ \bibinfo {pages} {115301} (\bibinfo {year} {2005})}\BibitemShut
  {NoStop}%
\bibitem [{\citenamefont {Gavrilov}\ \emph {et~al.}(2007)\citenamefont
  {Gavrilov}, \citenamefont {Gippius}, \citenamefont {Kulakovskii},\ and\
  \citenamefont {Tikhodeev}}]{Gavrilov2007.en}%
  \BibitemOpen
  \bibfield  {author} {\bibinfo {author} {\bibfnamefont {S.~S.}\ \bibnamefont
  {Gavrilov}}, \bibinfo {author} {\bibfnamefont {N.~A.}\ \bibnamefont
  {Gippius}}, \bibinfo {author} {\bibfnamefont {V.~D.}\ \bibnamefont
  {Kulakovskii}},\ and\ \bibinfo {author} {\bibfnamefont {S.~G.}\ \bibnamefont
  {Tikhodeev}},\ }\href {https://doi.org/10.1134/S1063776107050056} {\bibfield
  {journal} {\bibinfo  {journal} {JETP}\ }\textbf {\bibinfo {volume} {104}},\
  \bibinfo {pages} {715} (\bibinfo {year} {2007})}\BibitemShut {NoStop}%
\bibitem [{\citenamefont {Demenev}\ \emph {et~al.}(2008)\citenamefont
  {Demenev}, \citenamefont {Shchekin}, \citenamefont {Larionov}, \citenamefont
  {Gavrilov}, \citenamefont {Kulakovskii}, \citenamefont {Gippius},\ and\
  \citenamefont {Tikhodeev}}]{Demenev2008}%
  \BibitemOpen
  \bibfield  {author} {\bibinfo {author} {\bibfnamefont {A.~A.}\ \bibnamefont
  {Demenev}}, \bibinfo {author} {\bibfnamefont {A.~A.}\ \bibnamefont
  {Shchekin}}, \bibinfo {author} {\bibfnamefont {A.~V.}\ \bibnamefont
  {Larionov}}, \bibinfo {author} {\bibfnamefont {S.~S.}\ \bibnamefont
  {Gavrilov}}, \bibinfo {author} {\bibfnamefont {V.~D.}\ \bibnamefont
  {Kulakovskii}}, \bibinfo {author} {\bibfnamefont {N.~A.}\ \bibnamefont
  {Gippius}},\ and\ \bibinfo {author} {\bibfnamefont {S.~G.}\ \bibnamefont
  {Tikhodeev}},\ }\href {https://doi.org/10.1103/PhysRevLett.101.136401}
  {\bibfield  {journal} {\bibinfo  {journal} {Phys. Rev. Lett.}\ }\textbf
  {\bibinfo {volume} {101}},\ \bibinfo {pages} {136401} (\bibinfo {year}
  {2008})}\BibitemShut {NoStop}%
\bibitem [{\citenamefont {Krizhanovskii}\ \emph {et~al.}(2008)\citenamefont
  {Krizhanovskii}, \citenamefont {Gavrilov}, \citenamefont {Love},
  \citenamefont {Sanvitto}, \citenamefont {Gippius}, \citenamefont {Tikhodeev},
  \citenamefont {Kulakovskii}, \citenamefont {Whittaker}, \citenamefont
  {Skolnick},\ and\ \citenamefont {Roberts}}]{Krizhanovskii2008}%
  \BibitemOpen
  \bibfield  {author} {\bibinfo {author} {\bibfnamefont {D.~N.}\ \bibnamefont
  {Krizhanovskii}}, \bibinfo {author} {\bibfnamefont {S.~S.}\ \bibnamefont
  {Gavrilov}}, \bibinfo {author} {\bibfnamefont {A.~P.~D.}\ \bibnamefont
  {Love}}, \bibinfo {author} {\bibfnamefont {D.}~\bibnamefont {Sanvitto}},
  \bibinfo {author} {\bibfnamefont {N.~A.}\ \bibnamefont {Gippius}}, \bibinfo
  {author} {\bibfnamefont {S.~G.}\ \bibnamefont {Tikhodeev}}, \bibinfo {author}
  {\bibfnamefont {V.~D.}\ \bibnamefont {Kulakovskii}}, \bibinfo {author}
  {\bibfnamefont {D.~M.}\ \bibnamefont {Whittaker}}, \bibinfo {author}
  {\bibfnamefont {M.~S.}\ \bibnamefont {Skolnick}},\ and\ \bibinfo {author}
  {\bibfnamefont {J.~S.}\ \bibnamefont {Roberts}},\ }\href
  {https://doi.org/10.1103/PhysRevB.77.115336} {\bibfield  {journal} {\bibinfo
  {journal} {Phys. Rev. B}\ }\textbf {\bibinfo {volume} {77}},\ \bibinfo
  {pages} {115336} (\bibinfo {year} {2008})}\BibitemShut {NoStop}%
\bibitem [{\citenamefont {Wouters}\ and\ \citenamefont
  {Carusotto}(2007)}]{Wouters2007.prb.threshold}%
  \BibitemOpen
  \bibfield  {author} {\bibinfo {author} {\bibfnamefont {M.}~\bibnamefont
  {Wouters}}\ and\ \bibinfo {author} {\bibfnamefont {I.}~\bibnamefont
  {Carusotto}},\ }\href {https://doi.org/10.1103/PhysRevB.75.075332} {\bibfield
   {journal} {\bibinfo  {journal} {Phys. Rev. B}\ }\textbf {\bibinfo {volume}
  {75}},\ \bibinfo {pages} {075332} (\bibinfo {year} {2007})}\BibitemShut
  {NoStop}%
\bibitem [{\citenamefont {Dunnett}\ \emph {et~al.}(2018)\citenamefont
  {Dunnett}, \citenamefont {Ferrier}, \citenamefont {Zamora}, \citenamefont
  {Dagvadorj},\ and\ \citenamefont {{Szyma\ifmmode \acute{n}\else
  {\'n}\fi{}ska}}}]{Dunnett2018}%
  \BibitemOpen
  \bibfield  {author} {\bibinfo {author} {\bibfnamefont {K.}~\bibnamefont
  {Dunnett}}, \bibinfo {author} {\bibfnamefont {A.}~\bibnamefont {Ferrier}},
  \bibinfo {author} {\bibfnamefont {A.}~\bibnamefont {Zamora}}, \bibinfo
  {author} {\bibfnamefont {G.}~\bibnamefont {Dagvadorj}},\ and\ \bibinfo
  {author} {\bibfnamefont {M.~H.}\ \bibnamefont {{Szyma\ifmmode \acute{n}\else
  {\'n}\fi{}ska}}},\ }\href {https://doi.org/10.1103/PhysRevB.98.165307}
  {\bibfield  {journal} {\bibinfo  {journal} {Phys. Rev. B}\ }\textbf {\bibinfo
  {volume} {98}},\ \bibinfo {pages} {165307} (\bibinfo {year}
  {2018})}\BibitemShut {NoStop}%
\bibitem [{\citenamefont {Gavrilov}(2014)}]{Gavrilov2014.prb.b}%
  \BibitemOpen
  \bibfield  {author} {\bibinfo {author} {\bibfnamefont {S.~S.}\ \bibnamefont
  {Gavrilov}},\ }\href {https://doi.org/10.1103/PhysRevB.90.205303} {\bibfield
  {journal} {\bibinfo  {journal} {Phys. Rev. B}\ }\textbf {\bibinfo {volume}
  {90}},\ \bibinfo {pages} {205303} (\bibinfo {year} {2014})}\BibitemShut
  {NoStop}%
\bibitem [{\citenamefont {Gavrilov}\ \emph {et~al.}(2015)\citenamefont
  {Gavrilov}, \citenamefont {Brichkin}, \citenamefont {Grishina}, \citenamefont
  {Schneider}, \citenamefont {H{\"o}fling},\ and\ \citenamefont
  {Kulakovskii}}]{Gavrilov2015}%
  \BibitemOpen
  \bibfield  {author} {\bibinfo {author} {\bibfnamefont {S.~S.}\ \bibnamefont
  {Gavrilov}}, \bibinfo {author} {\bibfnamefont {A.~S.}\ \bibnamefont
  {Brichkin}}, \bibinfo {author} {\bibfnamefont {Y.~V.}\ \bibnamefont
  {Grishina}}, \bibinfo {author} {\bibfnamefont {C.}~\bibnamefont {Schneider}},
  \bibinfo {author} {\bibfnamefont {S.}~\bibnamefont {H{\"o}fling}},\ and\
  \bibinfo {author} {\bibfnamefont {V.~D.}\ \bibnamefont {Kulakovskii}},\
  }\href {https://doi.org/10.1103/PhysRevB.92.205312} {\bibfield  {journal}
  {\bibinfo  {journal} {Phys. Rev. B}\ }\textbf {\bibinfo {volume} {92}},\
  \bibinfo {pages} {205312} (\bibinfo {year} {2015})}\BibitemShut {NoStop}%
\bibitem [{\citenamefont {Whittaker}\ \emph {et~al.}(2017)\citenamefont
  {Whittaker}, \citenamefont {Dzurnak}, \citenamefont {Egorov}, \citenamefont
  {Buonaiuto}, \citenamefont {Walker}, \citenamefont {Cancellieri},
  \citenamefont {Whittaker}, \citenamefont {Clarke}, \citenamefont {Gavrilov},
  \citenamefont {Skolnick},\ and\ \citenamefont
  {Krizhanovskii}}]{Whittaker2017}%
  \BibitemOpen
  \bibfield  {author} {\bibinfo {author} {\bibfnamefont {C.~E.}\ \bibnamefont
  {Whittaker}}, \bibinfo {author} {\bibfnamefont {B.}~\bibnamefont {Dzurnak}},
  \bibinfo {author} {\bibfnamefont {O.~A.}\ \bibnamefont {Egorov}}, \bibinfo
  {author} {\bibfnamefont {G.}~\bibnamefont {Buonaiuto}}, \bibinfo {author}
  {\bibfnamefont {P.~M.}\ \bibnamefont {Walker}}, \bibinfo {author}
  {\bibfnamefont {E.}~\bibnamefont {Cancellieri}}, \bibinfo {author}
  {\bibfnamefont {D.~M.}\ \bibnamefont {Whittaker}}, \bibinfo {author}
  {\bibfnamefont {E.}~\bibnamefont {Clarke}}, \bibinfo {author} {\bibfnamefont
  {S.~S.}\ \bibnamefont {Gavrilov}}, \bibinfo {author} {\bibfnamefont {M.~S.}\
  \bibnamefont {Skolnick}},\ and\ \bibinfo {author} {\bibfnamefont {D.~N.}\
  \bibnamefont {Krizhanovskii}},\ }\href
  {https://doi.org/10.1103/PhysRevX.7.031033} {\bibfield  {journal} {\bibinfo
  {journal} {Phys. Rev. X}\ }\textbf {\bibinfo {volume} {7}},\ \bibinfo {pages}
  {031033} (\bibinfo {year} {2017})}\BibitemShut {NoStop}%
\bibitem [{\citenamefont {Sun}\ \emph {et~al.}(2012)\citenamefont {Sun},
  \citenamefont {Jia}, \citenamefont {Barsi}, \citenamefont {Rica},
  \citenamefont {Picozzi},\ and\ \citenamefont {Fleischer}}]{Sun2012}%
  \BibitemOpen
  \bibfield  {author} {\bibinfo {author} {\bibfnamefont {C.}~\bibnamefont
  {Sun}}, \bibinfo {author} {\bibfnamefont {S.}~\bibnamefont {Jia}}, \bibinfo
  {author} {\bibfnamefont {C.}~\bibnamefont {Barsi}}, \bibinfo {author}
  {\bibfnamefont {S.}~\bibnamefont {Rica}}, \bibinfo {author} {\bibfnamefont
  {A.}~\bibnamefont {Picozzi}},\ and\ \bibinfo {author} {\bibfnamefont {J.~W.}\
  \bibnamefont {Fleischer}},\ }\href {https://doi.org/10.1038/nphys2278}
  {\bibfield  {journal} {\bibinfo  {journal} {Nat. Phys.}\ }\textbf {\bibinfo
  {volume} {8}},\ \bibinfo {pages} {470} (\bibinfo {year} {2012})}\BibitemShut
  {NoStop}%
\bibitem [{\citenamefont {Gavrilov}(2016)}]{Gavrilov2016}%
  \BibitemOpen
  \bibfield  {author} {\bibinfo {author} {\bibfnamefont {S.~S.}\ \bibnamefont
  {Gavrilov}},\ }\href {https://doi.org/10.1103/PhysRevB.94.195310} {\bibfield
  {journal} {\bibinfo  {journal} {Phys. Rev. B}\ }\textbf {\bibinfo {volume}
  {94}},\ \bibinfo {pages} {195310} (\bibinfo {year} {2016})}\BibitemShut
  {NoStop}%
\bibitem [{\citenamefont {Leblanc}\ \emph {et~al.}(2020)\citenamefont
  {Leblanc}, \citenamefont {Malpuech},\ and\ \citenamefont
  {Solnyshkov}}]{Leblanc2020}%
  \BibitemOpen
  \bibfield  {author} {\bibinfo {author} {\bibfnamefont {C.}~\bibnamefont
  {Leblanc}}, \bibinfo {author} {\bibfnamefont {G.}~\bibnamefont {Malpuech}},\
  and\ \bibinfo {author} {\bibfnamefont {D.~D.}\ \bibnamefont {Solnyshkov}},\
  }\href {https://doi.org/10.1103/PhysRevB.101.115418} {\bibfield  {journal}
  {\bibinfo  {journal} {Phys. Rev. B}\ }\textbf {\bibinfo {volume} {101}},\
  \bibinfo {pages} {115418} (\bibinfo {year} {2020})}\BibitemShut {NoStop}%
\bibitem [{\citenamefont {Elesin}\ and\ \citenamefont
  {Kopaev}(1973)}]{Elesin1973}%
  \BibitemOpen
  \bibfield  {author} {\bibinfo {author} {\bibfnamefont {V.~F.}\ \bibnamefont
  {Elesin}}\ and\ \bibinfo {author} {\bibfnamefont {Y.~V.}\ \bibnamefont
  {Kopaev}},\ }\href {http://www.jetp.ac.ru/cgi-bin/e/index/e/36/4/p767?a=list}
  {\bibfield  {journal} {\bibinfo  {journal} {Sov. Phys. JETP}\ }\textbf
  {\bibinfo {volume} {36}},\ \bibinfo {pages} {767} (\bibinfo {year}
  {1973})}\BibitemShut {NoStop}%
\bibitem [{\citenamefont {Baas}\ \emph {et~al.}(2004)\citenamefont {Baas},
  \citenamefont {Karr}, \citenamefont {Eleuch},\ and\ \citenamefont
  {Giacobino}}]{Baas2004.pra}%
  \BibitemOpen
  \bibfield  {author} {\bibinfo {author} {\bibfnamefont {A.}~\bibnamefont
  {Baas}}, \bibinfo {author} {\bibfnamefont {J.~P.}\ \bibnamefont {Karr}},
  \bibinfo {author} {\bibfnamefont {H.}~\bibnamefont {Eleuch}},\ and\ \bibinfo
  {author} {\bibfnamefont {E.}~\bibnamefont {Giacobino}},\ }\href
  {https://doi.org/10.1103/PhysRevA.69.023809} {\bibfield  {journal} {\bibinfo
  {journal} {Phys. Rev. A}\ }\textbf {\bibinfo {volume} {69}},\ \bibinfo
  {pages} {023809} (\bibinfo {year} {2004})}\BibitemShut {NoStop}%
\bibitem [{\citenamefont {Carusotto}\ and\ \citenamefont
  {Ciuti}(2004)}]{Carusotto2004}%
  \BibitemOpen
  \bibfield  {author} {\bibinfo {author} {\bibfnamefont {I.}~\bibnamefont
  {Carusotto}}\ and\ \bibinfo {author} {\bibfnamefont {C.}~\bibnamefont
  {Ciuti}},\ }\href {https://doi.org/10.1103/PhysRevLett.93.166401} {\bibfield
  {journal} {\bibinfo  {journal} {Phys. Rev. Lett.}\ }\textbf {\bibinfo
  {volume} {93}},\ \bibinfo {pages} {166401} (\bibinfo {year}
  {2004})}\BibitemShut {NoStop}%
\bibitem [{\citenamefont {Haken}(1975)}]{Haken1975.rev}%
  \BibitemOpen
  \bibfield  {author} {\bibinfo {author} {\bibfnamefont {H.}~\bibnamefont
  {Haken}},\ }\href {https://doi.org/10.1103/RevModPhys.47.67} {\bibfield
  {journal} {\bibinfo  {journal} {Rev. Mod. Phys.}\ }\textbf {\bibinfo {volume}
  {47}},\ \bibinfo {pages} {67} (\bibinfo {year} {1975})}\BibitemShut {NoStop}%
\bibitem [{\citenamefont {Landman}\ \emph {et~al.}(1988)\citenamefont
  {Landman}, \citenamefont {Papanicolaou}, \citenamefont {Sulem},\ and\
  \citenamefont {Sulem}}]{Landman1988}%
  \BibitemOpen
  \bibfield  {author} {\bibinfo {author} {\bibfnamefont {M.~J.}\ \bibnamefont
  {Landman}}, \bibinfo {author} {\bibfnamefont {G.~C.}\ \bibnamefont
  {Papanicolaou}}, \bibinfo {author} {\bibfnamefont {C.}~\bibnamefont
  {Sulem}},\ and\ \bibinfo {author} {\bibfnamefont {P.~L.}\ \bibnamefont
  {Sulem}},\ }\href {https://doi.org/10.1103/PhysRevA.38.3837} {\bibfield
  {journal} {\bibinfo  {journal} {Phys. Rev. A}\ }\textbf {\bibinfo {volume}
  {38}},\ \bibinfo {pages} {3837} (\bibinfo {year} {1988})}\BibitemShut
  {NoStop}%
\bibitem [{\citenamefont {Gavrilov}(2020{\natexlab{a}})}]{Gavrilov2020.usp.en}%
  \BibitemOpen
  \bibfield  {author} {\bibinfo {author} {\bibfnamefont {S.~S.}\ \bibnamefont
  {Gavrilov}},\ }\href {https://doi.org/10.3367/ufne.2019.04.038549} {\bibfield
   {journal} {\bibinfo  {journal} {Phys. Usp.}\ }\textbf {\bibinfo {volume}
  {63}},\ \bibinfo {pages} {123} (\bibinfo {year}
  {2020}{\natexlab{a}})}\BibitemShut {NoStop}%
\bibitem [{Note1()}]{Note1}%
  \BibitemOpen
  \bibinfo {note} {The supplemental video file, which can be
    downloaded from the arXiv abstract page of this article, shows the
    real- and momentum-space dynamics of the system represented in
    Figs.~\ref{fig:dynamics}--\ref{fig:space}.}\BibitemShut {Stop}%
\bibitem [{\citenamefont {Cross}\ and\ \citenamefont
  {Hohenberg}(1993)}]{Cross1993}%
  \BibitemOpen
  \bibfield  {author} {\bibinfo {author} {\bibfnamefont {M.~C.}\ \bibnamefont
  {Cross}}\ and\ \bibinfo {author} {\bibfnamefont {P.~C.}\ \bibnamefont
  {Hohenberg}},\ }\href {https://doi.org/10.1103/RevModPhys.65.851} {\bibfield
  {journal} {\bibinfo  {journal} {Rev. Mod. Phys.}\ }\textbf {\bibinfo {volume}
  {65}},\ \bibinfo {pages} {851} (\bibinfo {year} {1993})}\BibitemShut
  {NoStop}%
\bibitem [{\citenamefont {Staliunas}\ and\ \citenamefont
  {S{\'a}nchez-Morcillo}(2003)}]{Staliunas.book}%
  \BibitemOpen
  \bibfield  {author} {\bibinfo {author} {\bibfnamefont {K.}~\bibnamefont
  {Staliunas}}\ and\ \bibinfo {author} {\bibfnamefont {V.~J.}\ \bibnamefont
  {S{\'a}nchez-Morcillo}},\ }\href {https://doi.org/10.1007/3-540-36416-1}
  {\emph {\bibinfo {title} {Transverse Patterns in Nonlinear Optical
  Resonators}}}\ (\bibinfo  {publisher} {Springer},\ \bibinfo {year}
  {2003})\BibitemShut {NoStop}%
\bibitem [{\citenamefont {Ferr{\'e}}\ \emph {et~al.}(2017)\citenamefont
  {Ferr{\'e}}, \citenamefont {Clerc}, \citenamefont {Coulibally}, \citenamefont
  {Rojas},\ and\ \citenamefont {Tlidi}}]{Ferre2017}%
  \BibitemOpen
  \bibfield  {author} {\bibinfo {author} {\bibfnamefont {M.~A.}\ \bibnamefont
  {Ferr{\'e}}}, \bibinfo {author} {\bibfnamefont {M.~G.}\ \bibnamefont
  {Clerc}}, \bibinfo {author} {\bibfnamefont {S.}~\bibnamefont {Coulibally}},
  \bibinfo {author} {\bibfnamefont {R.~G.}\ \bibnamefont {Rojas}},\ and\
  \bibinfo {author} {\bibfnamefont {M.}~\bibnamefont {Tlidi}},\ }\href
  {https://doi.org/10.1140/epjd/e2017-80072-3} {\bibfield  {journal} {\bibinfo
  {journal} {Eur. Phys. J. D}\ }\textbf {\bibinfo {volume} {71}},\ \bibinfo
  {pages} {172} (\bibinfo {year} {2017})}\BibitemShut {NoStop}%
\bibitem [{\citenamefont {Sarchi}\ \emph {et~al.}(2008)\citenamefont {Sarchi},
  \citenamefont {Carusotto}, \citenamefont {Wouters},\ and\ \citenamefont
  {Savona}}]{Sarchi2008}%
  \BibitemOpen
  \bibfield  {author} {\bibinfo {author} {\bibfnamefont {D.}~\bibnamefont
  {Sarchi}}, \bibinfo {author} {\bibfnamefont {I.}~\bibnamefont {Carusotto}},
  \bibinfo {author} {\bibfnamefont {M.}~\bibnamefont {Wouters}},\ and\ \bibinfo
  {author} {\bibfnamefont {V.}~\bibnamefont {Savona}},\ }\href
  {https://doi.org/10.1103/PhysRevB.77.125324} {\bibfield  {journal} {\bibinfo
  {journal} {Phys. Rev. B}\ }\textbf {\bibinfo {volume} {77}},\ \bibinfo
  {pages} {125324} (\bibinfo {year} {2008})}\BibitemShut {NoStop}%
\bibitem [{\citenamefont {Solnyshkov}\ \emph {et~al.}(2009)\citenamefont
  {Solnyshkov}, \citenamefont {Johne}, \citenamefont {Shelykh},\ and\
  \citenamefont {Malpuech}}]{Solnyshkov2009.j}%
  \BibitemOpen
  \bibfield  {author} {\bibinfo {author} {\bibfnamefont {D.~D.}\ \bibnamefont
  {Solnyshkov}}, \bibinfo {author} {\bibfnamefont {R.}~\bibnamefont {Johne}},
  \bibinfo {author} {\bibfnamefont {I.~A.}\ \bibnamefont {Shelykh}},\ and\
  \bibinfo {author} {\bibfnamefont {G.}~\bibnamefont {Malpuech}},\ }\href
  {https://doi.org/10.1103/PhysRevB.80.235303} {\bibfield  {journal} {\bibinfo
  {journal} {Phys. Rev. B}\ }\textbf {\bibinfo {volume} {80}},\ \bibinfo
  {pages} {235303} (\bibinfo {year} {2009})}\BibitemShut {NoStop}%
\bibitem [{\citenamefont {Gavrilov}(2018)}]{Gavrilov2018}%
  \BibitemOpen
  \bibfield  {author} {\bibinfo {author} {\bibfnamefont {S.~S.}\ \bibnamefont
  {Gavrilov}},\ }\href {https://doi.org/10.1103/PhysRevLett.120.033901}
  {\bibfield  {journal} {\bibinfo  {journal} {Phys. Rev. Lett.}\ }\textbf
  {\bibinfo {volume} {120}},\ \bibinfo {pages} {033901} (\bibinfo {year}
  {2018})}\BibitemShut {NoStop}%
\bibitem [{\citenamefont {Gavrilov}(2020{\natexlab{b}})}]{Gavrilov2020}%
  \BibitemOpen
  \bibfield  {author} {\bibinfo {author} {\bibfnamefont {S.~S.}\ \bibnamefont
  {Gavrilov}},\ }\href {https://doi.org/10.1103/PhysRevB.102.104307} {\bibfield
   {journal} {\bibinfo  {journal} {Phys. Rev. B}\ }\textbf {\bibinfo {volume}
  {102}},\ \bibinfo {pages} {104307} (\bibinfo {year}
  {2020}{\natexlab{b}})}\BibitemShut {NoStop}%
\bibitem [{\citenamefont {Gavrilov}(2017)}]{Gavrilov2017.en}%
  \BibitemOpen
  \bibfield  {author} {\bibinfo {author} {\bibfnamefont {S.~S.}\ \bibnamefont
  {Gavrilov}},\ }\href {https://doi.org/10.1134/S0021364017030079} {\bibfield
  {journal} {\bibinfo  {journal} {JETP Lett.}\ }\textbf {\bibinfo {volume}
  {105}},\ \bibinfo {pages} {200} (\bibinfo {year} {2017})}\BibitemShut
  {NoStop}%
\bibitem [{\citenamefont {Shelykh}\ \emph {et~al.}(2010)\citenamefont
  {Shelykh}, \citenamefont {Kavokin}, \citenamefont {Rubo}, \citenamefont
  {Liew},\ and\ \citenamefont {Malpuech}}]{Shelykh2010}%
  \BibitemOpen
  \bibfield  {author} {\bibinfo {author} {\bibfnamefont {I.~A.}\ \bibnamefont
  {Shelykh}}, \bibinfo {author} {\bibfnamefont {A.~V.}\ \bibnamefont
  {Kavokin}}, \bibinfo {author} {\bibfnamefont {Y.~G.}\ \bibnamefont {Rubo}},
  \bibinfo {author} {\bibfnamefont {T.~C.~H.}\ \bibnamefont {Liew}},\ and\
  \bibinfo {author} {\bibfnamefont {G.}~\bibnamefont {Malpuech}},\ }\href
  {https://doi.org/10.1088/0268-1242/25/1/013001} {\bibfield  {journal}
  {\bibinfo  {journal} {Semicond. Sci. Technol.}\ }\textbf {\bibinfo {volume}
  {25}},\ \bibinfo {pages} {013001} (\bibinfo {year} {2010})}\BibitemShut
  {NoStop}%
\bibitem [{\citenamefont {Vladimirova}\ \emph {et~al.}(2010)\citenamefont
  {Vladimirova}, \citenamefont {Cronenberger}, \citenamefont {Scalbert},
  \citenamefont {Kavokin}, \citenamefont {Miard}, \citenamefont
  {Lema\^{\i}{}tre}, \citenamefont {Bloch}, \citenamefont {Solnyshkov},
  \citenamefont {Malpuech},\ and\ \citenamefont {Kavokin}}]{Vladimirova2010}%
  \BibitemOpen
  \bibfield  {author} {\bibinfo {author} {\bibfnamefont {M.}~\bibnamefont
  {Vladimirova}}, \bibinfo {author} {\bibfnamefont {S.}~\bibnamefont
  {Cronenberger}}, \bibinfo {author} {\bibfnamefont {D.}~\bibnamefont
  {Scalbert}}, \bibinfo {author} {\bibfnamefont {K.~V.}\ \bibnamefont
  {Kavokin}}, \bibinfo {author} {\bibfnamefont {A.}~\bibnamefont {Miard}},
  \bibinfo {author} {\bibfnamefont {A.}~\bibnamefont {Lema\^{\i}{}tre}},
  \bibinfo {author} {\bibfnamefont {J.}~\bibnamefont {Bloch}}, \bibinfo
  {author} {\bibfnamefont {D.}~\bibnamefont {Solnyshkov}}, \bibinfo {author}
  {\bibfnamefont {G.}~\bibnamefont {Malpuech}},\ and\ \bibinfo {author}
  {\bibfnamefont {A.~V.}\ \bibnamefont {Kavokin}},\ }\href
  {https://doi.org/10.1103/PhysRevB.82.075301} {\bibfield  {journal} {\bibinfo
  {journal} {Phys. Rev. B}\ }\textbf {\bibinfo {volume} {82}},\ \bibinfo
  {pages} {075301} (\bibinfo {year} {2010})}\BibitemShut {NoStop}%
\bibitem [{\citenamefont {Gavrilov}\ \emph {et~al.}(2010)\citenamefont
  {Gavrilov}, \citenamefont {Brichkin}, \citenamefont {Dorodnyi}, \citenamefont
  {Tikhodeev}, \citenamefont {Gippius},\ and\ \citenamefont
  {Kulakovskii}}]{Gavrilov2010.jetpl.en}%
  \BibitemOpen
  \bibfield  {author} {\bibinfo {author} {\bibfnamefont {S.~S.}\ \bibnamefont
  {Gavrilov}}, \bibinfo {author} {\bibfnamefont {A.~S.}\ \bibnamefont
  {Brichkin}}, \bibinfo {author} {\bibfnamefont {A.~A.}\ \bibnamefont
  {Dorodnyi}}, \bibinfo {author} {\bibfnamefont {S.~G.}\ \bibnamefont
  {Tikhodeev}}, \bibinfo {author} {\bibfnamefont {N.~A.}\ \bibnamefont
  {Gippius}},\ and\ \bibinfo {author} {\bibfnamefont {V.~D.}\ \bibnamefont
  {Kulakovskii}},\ }\href {https://doi.org/10.1134/S0021364010150105}
  {\bibfield  {journal} {\bibinfo  {journal} {JETP Lett.}\ }\textbf {\bibinfo
  {volume} {92}},\ \bibinfo {pages} {171} (\bibinfo {year} {2010})}\BibitemShut
  {NoStop}%
\bibitem [{\citenamefont {Sekretenko}\ \emph {et~al.}(2013)\citenamefont
  {Sekretenko}, \citenamefont {Gavrilov},\ and\ \citenamefont
  {Kulakovskii}}]{Sekretenko2013.10ps}%
  \BibitemOpen
  \bibfield  {author} {\bibinfo {author} {\bibfnamefont {A.~V.}\ \bibnamefont
  {Sekretenko}}, \bibinfo {author} {\bibfnamefont {S.~S.}\ \bibnamefont
  {Gavrilov}},\ and\ \bibinfo {author} {\bibfnamefont {V.~D.}\ \bibnamefont
  {Kulakovskii}},\ }\href {https://doi.org/10.1103/PhysRevB.88.195302}
  {\bibfield  {journal} {\bibinfo  {journal} {Phys. Rev. B}\ }\textbf {\bibinfo
  {volume} {88}},\ \bibinfo {pages} {195302} (\bibinfo {year}
  {2013})}\BibitemShut {NoStop}%
\end{thebibliography}

\end{document}